\documentclass{ws-procs9x6}

\newcommand{\CO}{{\ensuremath \mathcal O}}
\newcommand{\msub}[1]{\ensuremath _{\mbox{\scriptsize #1}}} 

\begin{document}

\title{The status of pentaquark spectroscopy on the lattice\footnote{\uppercase{B}ased on
talks the authors gave at various conferences.}
}

\author{F. CSIKOR$^a$, Z. FODOR$^{a,b}$, 
        S.~D. KATZ$^b$\footnote{\uppercase{O}n leave from \uppercase{E}\"otv\"os
        \uppercase{L}or\'and \uppercase{U}niversity, \uppercase{B}udapest, 
        \uppercase{H}ungary.},  T.~G. KOV\'ACS$^c$}

\address{ $^A$Department of Theoretical Physics, \\
E\"otv\"os Lor\'and University, \\
H-1117 Budapest, \\
P\'azm\'any P\'eter s\'et\'any 1/a}

\address{ $^B$Department of Physics, \\
Bergische Universit\"at Wuppertal, \\
Gaussstr.\ 20 \\ 
D-42119 Germany}

\address{$^C$Department of Physics,\\
University of P\'ecs,\\
H-7624 P\'ecs, Ifj\'us\'ag u. 6 \\
Hungary}

\maketitle

\abstracts{The present work is a summary of the status of lattice 
pentaquark calculations. After a pedagogic introduction to 
the basics of lattice hadron spectroscopy we give a critical 
comparison of results presently available in the literature. 
Special emphasis is put on presenting some of the possible pitfalls 
of these calculations. In particular we discuss at length the choice of the 
hadronic operators and the separation of genuine five-quark states 
from meson-baryon scattering states.}

\section{Introduction}

The recent experimental searches for and the
discovery\cite{exp_theta,Aktas:2004qf} of the previously theoretically
predicted\cite{Diakonov:1997mm} exotic hadrons has
sparked considerable activity and gave rise
to diverse speculations regarding their structure, unexpectedly
small width, parity, isospin and spin. The only presently available
technique for computing low energy hadronic observables starting
from first principles (i.e.\ QCD) within systematically controllable 
approach is lattice QCD. 

All this said, it might seem surprising that of the more 
than 200 papers devoted to the subject of exotic baryons in the 
past year, there were only four lattice papers. Besides critically 
reviewing the currently available lattice results, in the present
work we also try to resolve this apparent paradox by discussing
some of the difficulties and pitfalls of the lattice approach.
The presentation is aimed for the general particle and nuclear
physics community. For this reason, in Section \ref{sec:Hsotl} we start
with an introduction to lattice hadron spectroscopy and also address
two points that are usually not discussed in great detail in lattice
papers, but are essential for the correct interpretation of lattice
pentaquark results.

In our opinion the biggest challenge lattice pentaquark calculations
face is how to choose the baryonic operators. Not only the
errors, but also the very possibility to identify  certain states 
depends crucially on the choice of operators. Unfortunately there is
very little guidance here and many technical restrictions.  
Subsection \ref{sec:Tcoo} is devoted to this issue.
Since the five-quark bound states we want to
study can be close to threshold, it is essential in any lattice spectroscopy
calculation to reliably distinguish between genuine
five-quark bound states and meson-baryon scattering states. In Subsection
\ref{sec:Stps} we discuss how this can be done. 

Having set the stage, in Section \ref{sec:R} we give a critical review of
the currently available lattice results and interpret them.
In Section \ref{sec:C} we conclude by summarizing the status of 
lattice calculations and stressing what is needed to be done for a final 
consolidation of the lattice results.

\section{Hadron spectroscopy on the lattice}
     \label{sec:Hsotl}

\subsection{The choice of operators}
     \label{sec:Tcoo}

In the framework of lattice QCD the role of the regulator is played by a
space-time lattice that replaces continuous space-time. As a result,
in a finite spatial volume the infinite dimensional 
functional integral turns into a mathematically 
well defined finite dimensional integral. The lattice also opens the 
way to the explicit numerical computation of hadronic observables by 
Euclidean Monte Carlo techniques. 

In hadron spectroscopy one would like to identify hadronic states with 
given quantum numbers. Practically this means the following.
We compute the vacuum expectation value of the Euclidean correlation function 
$\langle 0| \CO(t) \CO^\dagger(0) |0 \rangle$ of some composite hadronic
operator $\CO$. The operator $\CO$ is built 
out of quark creation and annihilation operators.
In physical terms the correlator is the amplitude 
of the ``process'' of creating a complicated hadronic state 
described by $\CO$ at time $0$ and destroying it at time $t$.

After inserting a complete set of eigenstates 
$|i \rangle$ of the full QCD Hamiltonian the correlation function
can be written as
\begin{equation}
  \langle 0| \CO(t) \CO^\dagger(0) |0 \rangle = 
  \sum_i \;\; |\; \langle i | \CO^\dagger(0) |0 \rangle \; |^2 \; 
   \; \mbox{e}^{-(E_i-E_0)t},
     \label{eq:corr}
\end{equation}
where 
\begin{equation}
  \CO(t) = \mbox{e}^{-Ht}\; \CO(0) \; \mbox{e}^{Ht}
\end{equation}
and $E_i$ are the energy eigenvalues of the Hamiltonian. 

Note that since we work in Euclidean space-time 
(the real time coordinate $t$ is replaced with $-it$), the
correlators do not oscillate, they rather die out exponentially
in imaginary time. In particular, after long enough time 
only the lowest (few) state(s) created by $\CO$ give 
contribution to the correlator. The energy eigenvalues 
corresponding to those states can be extracted from exponential
fits to the large $t$ behaviour of the correlator.

In the simplest cases one is typically interested in hadron masses.
A trivial but most important requirement in the choice of $\CO$ is
that it should have the quantum numbers of the state we intend to study.
Otherwise the overlap $\langle i | \CO^\dagger(0) |0 \rangle$ would be
zero and the corresponding exponent could not be extracted.
In order to have optimal overlap with only one state $|i\rangle$, 
$\CO^\dagger(0) |0 \rangle$ should be as ``close'' to
$|i\rangle$ as possible.

A hadron mass is the ground state energy in a sector with
given internal quantum numbers and zero momentum.
Projection to the zero momentum sector is achieved by summing a 
local operator over all of three-space as
\begin{equation}
 \CO(\vec{p}=0) = \sum_{\vec{x}} \mbox{e}^{-i\vec{p}\vec{x}} 
 \CO(0,\vec{x})|_{\vec{p}=0} = \sum_{\vec{x}} \CO(0,\vec{x}).
\end{equation}

 One of the most important experimentally still unknown quantum numbers
of pentaquark states is their parity. Thus, we also briefly touch upon
the parity assignment on the lattice. The simplest 
baryonic operators do not create parity eigenstates, rather they
couple to both parity channels. Projection to the $+/-$ parity 
eigenstates can be performed as 
\begin{equation}
 \CO_{\pm} = \frac{1}{2}(\CO \pm  P \CO P^{-1}).
\end{equation}
For the simplest operators the parity operator $P$ acts on $\CO$ as 
\begin{equation}
 P \CO P^{-1} = \eta \gamma_0 \CO,
     \label{eq:ip}
\end{equation}
where $\eta=\pm 1$ is the internal parity of the operator $\CO$. For
more complicated operators, in particular for non-pointlike ones, this
might become more involved. If the parity of a state is not known, 
it can be determined by computing 
the correlator in both parity channels and deciding which channel produces
a mass closer to the experimentally observed one.

All quantum numbers fixed, there is still considerable freedom in
the choice of $\CO$. This freedom has to be exploited to ensure 
maximal overlap of $\CO^\dagger(0) |0 \rangle$ with the desired state and minimal
overlap with close-by competing, but unwanted states. This is 
essential not only for smaller errors. With the wrong choice of
$\CO$ the desired state might be practically undetectably lost
in the noise.  Unfortunately, beyond the quantum numbers there 
is usually little if any guidance in the choice of $\CO$ and
herein lies the biggest challenge of lattice pentaquark spectroscopy.
It is almost impossible to {\em disprove} the existence of a given
state. If one cannot detect it with a given operator $\CO$ , it might
just mean that $\CO$ has too small overlap with the desired state and
the signal is lost in the noise. Indeed, even in the case of the
nucleon simple operators are known that have the correct quantum numbers,
but too little overlap with the nucleon ground state and no nucleon signal
can be extracted from their correlator\cite{Sasaki:2001nf}.

If the wave function of the quarks in the given hadronic state were 
known, that would dictate the form of the operator to be used. In the
case of pentaquarks there are several suggestions in
the literature and in principle it would be interesting to try operators
corresponding to at least some of them.
There are, however, two serious restrictions lattice calculations
face in this respect. The first one concerns the spatial structure of
the wave function, the second one its index structure. In the remainder
of this section we discuss these.

Concerning the spatial structure of the wave function, we have to note that 
the correlation function in Eq.\ (\ref{eq:corr}) is computed on the lattice
by decomposing it in terms of single quark correlators 
$\langle 0| q_\alpha(x) q_\beta^\dagger(y) |0 \rangle$. Those in
turn are simply the matrix elements $D^{-1}(x,\alpha;y,\beta)$
of the inverse of the lattice Dirac operator. If 
$\CO$ were to be based on an arbitrary five-quark 
wave function, the brute force computation of the correlator of $\CO$
would in general require
quark propagators $D^{-1}(x,\alpha;y,\beta)$ from any space-time point
$x$ to any other point $y$. On currently used lattice sizes this would
require the computation and storage of order $10^{13}$ matrix elements,
taking up about 100~Terabytes and requiring hundreds of Teraflops of
CPU power. This is clearly out of reach for presently available computers.

The only way around is to fix a quark wave function
$\psi_\beta(\vec{y})$ and store only the matrix elements 
\begin{equation}
d(x\alpha) = \sum_{\vec{y}\beta}D^{-1}(x,\alpha;y_0=0,\vec{y},\beta)\psi_\beta(\vec{y}).
\end{equation}
This choice drastically cuts down the computing requirements. Unfortunately, 
at the same time it also restricts $\CO$ to be built as a 
product of single quark wave functions with
the single quarks being in some state $\psi$. 
One needs to perform as many Dirac operator
inversions as the number of different quark wave functions contained in 
$\CO$. Since Dirac operator inversion is usually
the most expensive part of these
computations, one typically settles with using only two different quark
sources, one for the light quarks and one for the strange quark.
In fact, all four lattice pentaquark studies have used this simplest 
choice. 

Besides the spatial structure of $\CO$ the single quark spin, colour and 
flavour indices also have to be arranged properly for $\CO$ to have the
desired quantum numbers. Even then the arrangement of indices is also not unique. 
An additional difficulty one faces here compared to conventional three
quark hadron spectroscopy is that index summation becomes exponentially
more expensive if we increase the number of quarks. While with three quarks this 
part of the calculation is usually negligible, even for
the simplest five quark operators it takes up around 50\% of the
CPU time. This circumstance restricted the choice of pentaquark
operators so far to the simplest ones. 

To illustrate how these issues appear in practice we now discuss a few
specific examples of $\CO$ that have already been used. In the first
lattice study\cite{Csikor:2003ng} $\CO$ had the same Dirac structure as
that of nucleon plus kaon system, but colour indices were contracted 
differently, as\cite{Zhu:2003ba}
\begin{equation}
 \CO_{I=0/1} =
            \epsilon_{abc}\; [u_a^T C\gamma_5 d_b]\; 
           \{u_{e} \, \bar{s}_e i\gamma_5 d_{c} 
           \; \mp \; (u \leftrightarrow d) \}, 
\end{equation}
where $I=0/1$ and the two alternative signs correspond to the isospin
singlet and triplet channel respectively. One could also contract 
the colour indices as in the nucleon$\times$kaon, a choice used by
Mathur et al.\cite{Liu}. 

Another possible way to contract the quark indices in $\CO$ is according
to the diquark-diquark-antiquark picture of Jaffe and 
Wilczek\cite{Jaffe:2003sg}. They proposed to insert the two diquarks in  
\begin{equation}
 \CO_{I=0} = \epsilon_{adg} \; 
            [\epsilon_{abc}\; u_b^T C\gamma_5 d_c]\;\;  
            [\epsilon_{def}\; u_e^T C\gamma_5 d_f]\;\; 
                                        C \bar{s}_g^T.
     \label{eq:dda}
\end{equation}
in a relative P-wave. 

In general, in a diquark-diquark-antiquark wave function of the form
(\ref{eq:dda}) the two diquarks must be in different quantum 
states\footnote{Otherwise the operator identically vanishes due to its symmetry 
with respect to the interchange of the diquarks.}. 
On the lattice, that would require the computation of several quark propagators.
Instead, Sasaki avoided the diquark-diquark symmetry by omitting a 
$\gamma_5$ from one of the diquarks\cite{Sasaki:2003gi}. The operator
he, and following in  his footsteps subsequently Chiu \& Hsieh\cite{Chiu:2004gg} 
considered, was
\begin{equation}
 \CO_{I=0} = \epsilon_{adg} \; 
           [ \epsilon_{abc}\; u_b^T C d_c]\;\;  
           [ \epsilon_{def}\; u_e^T C\gamma_5 d_f]\;\; 
                                        C \bar{s}_g^T.
\end{equation}
     
In summary, both in terms of spatial and index structure there are many 
more possibilities for $\CO$, but on the lattice they all require considerably
more CPU time than the ones explored so far. However, we expect that several
other possibilities will be tried in the near future.

\subsection{Separating two particle states}
     \label{sec:Stps}

Pentaquark spectroscopy is further complicated by the presence 
of two-particle scattering states lying close to the 
pentaquark state. Lattice calculations are always performed in
a finite spatial volume, therefore these scattering states do
not form a continuum. They occur at discrete energy values dictated
by the discrete momenta $p_k = 2k\pi/L, k=0,1,...$, allowed in a 
box of linear size $L$. In lattice pentaquark computations it is
absolutely essential to be able to distinguish between these two-particle
nucleon-meson scattering states and genuine five quark bound states.

In fact, the first experimentally found exotic baryon state, the
$\Theta^+(1540)$ lies just about 100~MeV {\it above} the
nucleon-kaon threshold. This implies that for large enough time 
separation the correlation function is bound to be dominated by
the nucleon-kaon state. However, the mass difference between the two
states is quite small and the mass of the $\Theta^+$ might still be
reliably extracted in an intermediate time window, provided that
\begin{equation}
 |\langle \Theta^+ | \CO | 0 \rangle| \; \gg \; 
        | \langle N+K | \CO | 0 \rangle |.
\end{equation}
Even then, identifying the $\Theta^+$ is still a non-trivial matter
since the $\Theta^+$ ground state is embedded in an infinite tower of
nucleon kaon scattering states with relative momenta allowed by the 
finite spatial box. Since the parity of the $\Theta^+$ is unknown, we
have to consider both parity channels. The situation is qualitatively 
different in the two channels.

If the $\Theta^+$ had positive parity, its lattice identification
would be somewhat simpler. This is because due to the negative 
internal parity of the kaon it is only the scattering states with
odd angular momentum that produce positive parity. The scattering
state with zero relative linear momentum does not couple to these and
consequently it does not appear in the positive parity channel.
Therefore, the lowest scattering state here 
has relative momentum $p = 2\pi/L$ and it is above the
$\Theta^+$, provided the linear size of the spatial box is smaller
than 4.5~fm. The box can thus be chosen 
small enough to ensure that the $\Theta^+$ is the lowest state 
with positive parity and also to leave a large enough energy gap
for its safe identification.

The situation is much less favourable in the negative parity channel.
Using a similar argument one can show that here
it is always the $p\msub{rel}=0$ scattering state
that is the lowest. The best we can do is that with 
the proper choice of the spatial volume
the $\Theta^+$ ground state can be the second lowest state. 
Due care must be taken to ensure that $\Theta^+$ is between the first two 
scattering states, well separated from both of them. This is essential
because the reliable identification of higher lying states is much 
more difficult. 

Finally, for a convincing confirmation of the 
pentaquark state in either parity channel, one also has to
identify the competing scattering states observing the
volume dependence dictated by the allowed smallest momentum. This
would clearly require a finite volume analysis combined with
a reliable method to extract {\em several} low lying states from the
spectrum. Apart from the volume dependence of the masses,
another powerful tool to distinguish between two-particle
and one-particle states is to check the volume dependence of
their spectral weights\cite{Liu}.

There are essentially two possible ways of identifying more than one
low lying state from correlators. Firstly, if there is a time interval
where more than one state has an appreciable contribution to the
correlator, a sum of exponentials can also be fitted as
\begin{equation}
 \langle 0| \CO(t) \CO^\dagger(0) |0 \rangle = 
 C_1\mbox{e}^{-E_0t}+C_2\mbox{e}^{-E_1t}+...
\end{equation}
For this method to yield reliable energy estimates for higher states,
one usually needs extremely good quality data. 

The other possibility is to make use of several different operators,
compute all possible cross-correlators 
and diagonalize the Hamiltonian in the subspace spanned by the 
states created by those operators\cite{Luscher:1990ck,Sasaki:2001nf,Burch:2004he}. 
This is a very powerful method to identify excited states and it can
also be combined with the first possibility.

\subsection{Extrapolations, sources of errors and uncertainties}
     \label{se:Eaosoe}

\begin{figure}[ht]
\centerline{\epsfxsize=4.1in\epsfbox{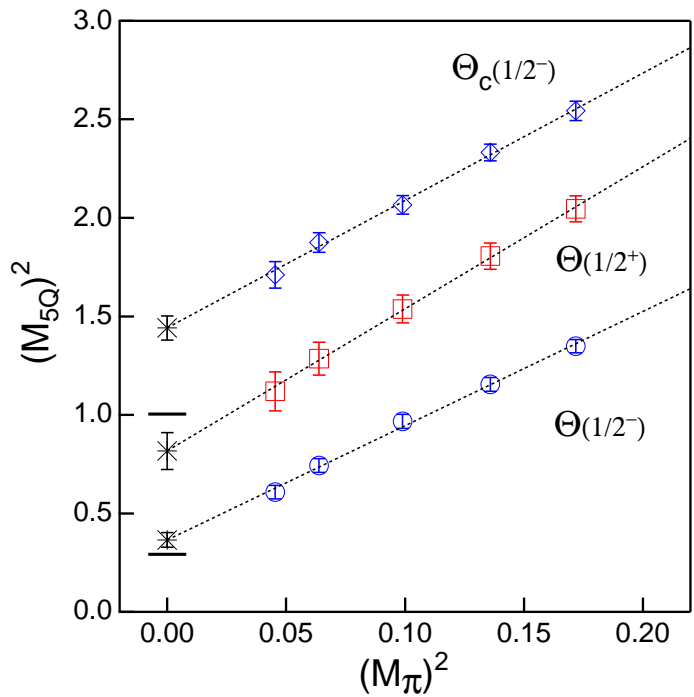}}   
\caption{Chiral extrapolation of the masses different five quark states
from Ref.$^7$
\label{fig:ThetaSqMass}}
\end{figure}

The lattice spectroscopy of hadrons built out of light quarks
involves two extrapolations. Firstly, simulations at the physical
$u/d$ quark masses would presently be prohibitively expensive,
therefore one has to do several calculations with heavier
quarks and then extrapolate to the physical quark masses.
A set of typical chiral extrapolations are shown in 
Fig.~\ref{fig:ThetaSqMass}.

The lightest quarks used in presently available pentaquark
studies correspond to pion masses in the range 180-650~MeV
(see Table~\ref{table}). 

\begin{table}[!ht]
\tbl{Lattice spacing and smallest pion mass of lattice
pentaquark calculations.}
{ 
\begin{tabular}{cccc} \hline
                       &  action & $a$ (fm)   &  smallest $m_\pi$ (MeV) \\
                 \hline \hline 
 Csikor et al. &  Wilson & 0.17-0.09  &  420 \\ \hline
 Sasaki        &  Wilson & 0.07       &  650 \\ \hline
 Liu et al.    &  chiral & 0.20       &  180 \\ \hline
 Chiu \& Hsieh   &  chiral & 0.09       &  400 \\ \hline
\end{tabular} \label{table} }
\vspace*{-13pt}
\end{table}

Secondly, the space-time lattice is not a physical entity, it is
just a regulator that has to be eventually removed to recover 
continuous space-time. This implies that physical quantities have
to be computed on lattices of different mesh sizes and extrapolated
to the zero lattice spacing (continuum) limit. Lattice simulations 
can differ from one another in many technical details and it is only
the continuum limit of physical quantities that is meaningful 
to compare among different simulations. 

In the remainder of this Subsection we briefly summarize the sources
of errors and uncertainties in lattice simulations indicating 
also how to handle them.

\begin{itemize}
\item {\em Statistical errors} are well understood and can be kept
at bay by increasing the statistics.
\item {\em Extrapolations} in quark mass and lattice spacing are another
source of uncertainty. Fortunately mass ratios of hadrons are usually 
quite insensitive in the present range of parameters.
\item{\em Quenching,} i.e.\ neglecting the fermion determinant
(omitting quark loops) is still a necessary compromise we have to
live with in most of the lattice calculations. Fortunately experience
tells us that stable hadron mass ratios have only a few per cent quenching
error.
\item{\em Finite volume} effects constitute another potential source
of error. There are different sources of volume dependence that can 
be properly accounted for and even be used to distinguish between
bound states and two particle scattering states.
\item As we have already discussed the desired state can be {\em contaminated
from other nearby states,} but this can be taken care of by a combination  
of the cross correlator technique and a careful finite volume analysis.
\item Finally there is a theoretical uncertainty originating
in the lack of any guidance in the {\em choice of operators} and
the inability to choose $\CO$ optimally. This can result in larger 
statistical errors or even in a complete failure to identify an
existing state. For this reason it is almost impossible to rule out
the existence of a state with given energy and quantum numbers.
\end{itemize}

\section{Results}
     \label{sec:R}

Having set the stage we can now present the lattice results along with
our interpretation. Four independent lattice pentaquark studies have 
been presented. Their main results can be summarized as
follows.
\begin{itemize}
\item {\em Csikor, Fodor, Katz and Kovacs}\cite{Csikor:2003ng} 
identified a state in the
 $I^P=0^-$ channel with a mass consistent with the experimental $\Theta^+$
 and the lowest mass found in the opposite parity $I^P=0^+$
 channel was significantly higher. Using $2\times 2$ cross correlators an 
 attempt was also made to separate the $\Theta^+$ and the lowest nucleon 
 kaon state.
\item {\em Sasaki}\cite{Sasaki:2003gi} using a different operator and 
double exponential fits, subsequently also found a state consistent with
the $\Theta^+$ also in the $I^P=0^-$ channel. He also managed to identify the 
charmed analogue of the $\Theta^+$ 640~MeV above the $DN$ threshold. (The 
experimentally found anticharmed pentaquark lies only about 
300~MeV above the threshold.)
\item {\em Liu et al.}\cite{Liu} reported that they could not
see any state compatible with the $\Theta^+$ in either parity isosinglet
channel. Although their smallest pion mass was the closest to the physical
one and the use an improved, chiral Dirac operator, they utilized 
the nucleon$\times$kaon operator and their lattice
is the coarsest of the four studies\footnote.
On the other hand they made use of sophisticated multi-exponential fits
with Bayesian priors.
\item Finally {\em Chiu~\&~Hsieh}\cite{Chiu:2004uh}, in disagreement
with the first two studies, saw a positive parity isosinglet
state compatible to the $\Theta^+$, whereas the lowest state they found
in the negative parity state was much higher. In a subsequent 
paper\cite{Chiu:2004uh} they also identified states claimed to be 
charmed counterparts of the $\Theta^+$.   
\end{itemize}

Our tentative interpretation of this somewhat controversial situation is 
as follows. Liu et al.\ used only one operator with exactly the same 
index structure as that of the nucleon kaon system. This might explain
why they see only the expected scattering states. 

The three remaining studies could be interpreted to have found genuine 
pentaquark states. All three agree that the lowest masses in the two 
parity channels differ by about 50\%, but they do not agree on the parity
of the $\Theta^+$ state. 
While Csikor et al.\ and Sasaki suggest negative parity, Chiu~\&~Hsieh
claim positive parity. According to the interpretation of Chiu~\&~Hsieh they
found different parity because they used a quark action with better behaviour
at small quark masses, albeit the same operator as Sasaki. 
The pion masses they use $(\geq 400$~MeV) overlaps with those
of Sasaki $(\geq 650$~MeV). In this region using the same hadron
operator all other hadron masses in the literature obtained with
these two quark actions agree (see e.g.\cite{Chiu:2004gg,Aoki:2002fd}). 
Thus it is extremely unlikely that the same operator with different 
lattice actions produces such vastly different masses.

In our opinion a more likely resolution
of this contradiction is that someone might have simply
misidentified the parity. 
On the one hand, the results of Chiu~\&~Hsieh and on the other hand, those
of Sasaki (and Csikor et al.) would become
compatible with each other if parities were flipped in one of them. 
A possible hint for a parity mismatch is provided by Chiu~\&~Hsieh in
their second paper\cite{Chiu:2004uh}. 
They considered two operators with opposite internal parities, but
otherwise having exactly the same quantum  numbers. 
Contrary to physical expectations, their ordering
of the lowest mass states in the two parity channels turned out
to depend on the internal parity of the operator. This suggests that
internal parity might not have been properly
taken into account (see Eq.\ \ref{eq:ip}).
Finally we would like to note that at this stage we can merely offer these
speculations and the issue has to be resolved by an independent study.

\section{Conclusions}
     \label{sec:C}

In summary, lattice QCD is the only known systematic approach to
calculate the features of the pentaquarks from first principles
(i.e.\ QCD). There have been four 
independent exploratory lattice pentaquark studies so
far with somewhat different findings. One of them sees only the expected
scattering state. Three analyses suggest mass states around the experimentally
detected pentaquarks. In order to justify these signals as 
pentaquark states one should convincingly separate them from 
the existing nearby scattering states. None of the
groups carried out this analysis. Furthermore, it should be 
realized that none of these analyses can be complete
for the following reason. In such a complete analysis one should see
the pentaquark in one parity channel and the lowest expected
scattering state in the other. All of the three groups reported energy
states coinciding with the pentaquark mass in one of the parity
channels; however, in the other channel the energy state is much higher
than the expected scattering state.

Since both parities have been suggested by lattice works, 
at least one of the results
will coincide with the parity to be found experimentally.
Nevertheless, no convincing final answer from
lattice QCD can be claimed unless the above program
has been completed. More specifically, it cannot be ruled out that
pentaquark states observed so far on the lattice turn out to be mixtures
of nucleon-kaon scattering states.

As we already emphasized, for a full picture one needs to systematically
map out the lowest few states in all interesting channels.
This will most likely be possible only with the use of
non-trivial spatial quark wave functions, the study of several operators
and the cross-correlator technique combined with a careful finite
volume analysis. This is currently under way and
we hope to be able to report new results in the near future.

\section*{Acknowledgments}
Partial support from 
OTKA Hungarian science grants under contract No.
T034980/\-T037615/\-TS044839/\-T046925 is acknowledged. 
T.G.K. was also supported through a Bolyai Fellowship.

\end{document}